# Unidentified Infrared Emission Features in Mid-infrared Spectrum of Comet 21P/Giacobini-Zinner

Takafumi Ootsubo[1], Hideyo Kawakita[2,3], Yoshiharu Shinnaka[2], Jun-ichi Watanabe[4], Mitsuhiko Honda[5]

[1] Institute of Space and Astronautical Science (ISAS), Japan Aerospace Exploration Agency (JAXA), 3-1-1, Yoshinodai, Chuo-ku, Sagamihara, Kanagawa 252-5210, Japan

[2] Laboratory of Infrared High-resolution Spectroscopy (LiH), Koyama Astronomical Observatory, Kyoto Sangyo University, Motoyama, Kamigamo, Kita-ku, Kyoto 603-8555, Japan

[3] Department of Astrophysics and Atmospheric Sciences, Faculty of Science, Kyoto Sangyo University, Motoyama, Kamigamo, Kita-ku, Kyoto 603-8555, Japan

[4] Public Relation Center, National Astronomical Observatory of Japan, 2-21-1 Osawa, Mitaka, Tokyo 181-8588, Japan

[5] Faculty of Biosphere-Geosphere Science Department of Biosphere-Geosphere Science, Okayama University of Science, 1-1 Ridaicho, Kita-ku, Okayama-shi 700-0005, Japan

**Corresponding author:**
Takafumi Ootsubo
Japan Aerospace Exploration Agency, Institute of Space and Astronautical Science
3-1-1 Yoshinodai, Chuo-ku, Sagamihara, Kanagawa 252-5210, Japan
Phone: +81-50-336-23179
Fax: +81-42-786-7202
E-mail: ootsubo@ir.isas.jaxa.jp

ABSTRACT

Comet 21P/Giacobini-Zinner (hereafter, comet 21P/G-Z) is a Jupiter-family comet and a parent comet of the October Draconids meteor shower. If meteoroids originating from a Jupiter-family comet contain complex organic molecules, such as amino acids, they are essential pieces of the puzzle regarding the origin of life on Earth. We observed comet 21P/G-Z in the mid-infrared wavelength region using the Cooled Mid-infrared Camera and Spectrometer (COMICS) on the 8.2 m Subaru Telescope on UT 2005 July 5. Here, we report the unidentified infrared (UIR) emission features of comet 21P/G-Z, which are likely due to complex organic molecules (both aliphatic and aromatic hydrocarbons), and the thermal emission from amorphous/crystalline silicates and amorphous carbon grains in its mid-infrared low-resolution spectrum. The UIR features at ~8.2 μm, ~8.5 μm, and ~11.2 μm found in the spectrum of comet 21P/G-Z could be attributed to polycyclic aromatic hydrocarbons (or hydrogenated amorphous carbons) contaminated by N- or O-atoms, although part of the feature at ~11.2 μm comes from crystalline olivine. The other feature at ~9.2 μm might originate from aliphatic hydrocarbons. Comet 21P/G-Z is enriched in complex organic molecules. Considering that the derived mass fraction of crystalline silicates in comet 21P/G-Z is typical of comets, we propose that the comet originated from a circumplanetary disk of giant planets (similar to Jupiter and Saturn) where was warmer than the typical comet-forming region (5–30 au from the Sun) and was suitable for the formation of complex organic molecules. Comets from circumplanetary disks might be enriched in complex organic molecules, such as comet 21P/G-Z, and may have provided pre-biotic molecules to ancient Earth by direct impact or meteor showers.

1. Introduction

Comet 21P/Giacobini-Zinner (hereafter, comet 21P/G-Z) was discovered in 1900 by Michel Giacobini and independently rediscovered by Ernst Zinner in 1913. It is a Jupiter-family comet with an orbital period of ~6.6 years in a low inclination orbit (its *Tisserand* parameter is $T_J$ = 2.47). Comet 21P/G-Z is considered a parent body of the October Draconids meteor shower based on the similarity of the orbits of the comet and its meteoroids (Jenniskens 2006). This comet is peculiar in terms of both its volatile content and dust grains in contrast to other comets; i.e., (i) it is depleted in carbon-chain molecules, $NH_2$, and highly volatile species, and (ii) has a negative linear polarization gradient produced by grains.

Based on the optical spectrophotometric survey of comets (A'Hearn et al. 1995), comet 21P/G-Z is depleted in $C_2$, $C_3$, and NH compared to CN. The low-resolution optical spectroscopic survey revealed that comet 21P/G-Z is depleted in both $C_2$ and $NH_2$, but normal for CN, compared to $H_2O$ (Fink 2009). Fink (2009) proposed four taxonomic classes based on the survey and comet 21P/G-Z is a prototype of "G-Z type" (~6 % of comets in the survey). It is also reported that comet 21P/G-Z is depleted in highly volatile species ($C_2H_6$, $CH_3OH$, and CO) compared to Oort Cloud comets (e.g., Weaver et al. 1999; DiSanti et al. 2013; Dello Russo et al. 2016). The comet is normal for CO but depleted in $C_2H_6$ and $CH_3OH$ in comparison with a typical Jupiter-family comet (Dello Russo et al. 2016). Thus, comet 21P/G-Z seems to be peculiar even as a Jupiter-family comet from the viewpoint of the chemistry of cometary volatiles.

Furthermore, it is reported that the linear polarization for dust continuum in optical wavelength region is lower for longer wavelength region in comet 21P/G-Z (usually higher for longer wavelength region in other comets), and suggested that the negative wavelength gradient of polarization seen in comet 21P/G-Z might be explained by a higher content of organic materials in dust grains, a high abundance of large particles, or a combination of both (Kiselev et al. 2000a, 2000b). To date, only some comets are known to show such a negative wavelength gradient of polarization (Kiselev et al. 2015). However, complex high-molecular-weight organic molecules like polycyclic aromatic hydrocarbons (PAHs) have never been detected clearly in comet 21P/G-Z and also in other comets, except in comet 67P/Churyumov–Gerasimenko by *in situ* observations of Rosetta spacecraft (Goesmann et al. 2015; Fray et al. 2016; Altwegg et al. 2019).

Comets having low inclination angles (such as Jupiter-family comets) and large dust particles released from them can collide with the Earth more easily than the Oort Cloud comets. If the complex organic molecules like amino acids are enriched in comet 21P/G-Z and in the meteoroids of the October Draconids, the comets having a similar origin as comet 21P/G-Z and their meteor showers might deliver complex organic materials to the ancient Earth.

2. Observations and Data Reduction

We conducted mid-infrared spectroscopic and imaging observations of comet 21P/G-Z on UT 2005 July 5 (the perihelion passage of the comet was on UT 2005 July 2.75). The heliocentric and geocentric distances of the comet were 1.04 and 1.43 au, respectively. Both imaging observations with narrow-band filters (*N*8.8, *N*12.4, and *Q*18.8, which are centered at $\lambda$ = 8.8, 12.4, and 18.8 μm with $\Delta\lambda$ = 0.8, 1.2, and 0.9 μm, respectively) and N-band (*NL spec*; $\lambda$ = 8–13 μm) low-resolution spectroscopic observations ($\lambda/\Delta\lambda \sim 250$) were performed with the Cooled Mid-infrared Camera and Spectrometer (COMICS) on the 8.2 m Subaru Telescope on Mauna Kea, Hawaii (Kataza et al. 2000; Okamoto et al. 2003). We used a slit of 0.33 arcsec in width and 40 arcsec in length for the low-resolution spectroscopy. To cancel high-background radiation in the N-band observations, secondary-mirror-chopping was used at a frequency of 0.45–0.43 Hz with an amplitude of 15 arcsec; i.e., the target was placed on the 40-arcsec-length slit at two positions. We also observed the standard star (HD3712) chosen by Cohen et al. (1999) in addition to the comet. The flux calibration and correction of atmospheric absorption were performed based on the photometric observations and spectra of the standard star. The template spectrum for the standard star was provided for the absolute flux calibration by Cohen et al. (1999). The wavelength calibration was performed based on the sky emission lines. The spectroscopic and photometric data taken by COMICS were reduced using IRAF[1] software and the special tools provided by the COMICS instrument team[2]. The aperture size for the photometry was set to 2.6 arcsec in radius. The spectra were extracted with a slit region of 0.33 arcsec (slit width) by 1.65 arcsec (along the slit direction) to achieve the adequate signal-to-noise ratio of the extracted spectrum. Slit-loss corrections for the spectra were adopted based on the photometric data. The details of the observation are listed in Table 1.

---

[1] IRAF is distributed by the National Optical Astronomy Observatory, which is operated by the Association of Universities for Research in Astronomy (AURA) under cooperative agreement with the National Science Foundation.

[2] https://subarutelescope.org/Observing/DataReduction/index.html

3. Results

The obtained spectrum is shown in Figure 1. The spectrum of comet 21P/G-Z has distinct emission peaks at approximately 8.5, 9.2, and 11.2 μm and weak peaks at approximately 8.2, 11.6, and 11.9 μm. The mid-infrared spectra of comets are usually dominated by emissions from silicate and carbonaceous materials. Silicate grains are Mg-Fe olivine and pyroxene in both amorphous and crystalline forms. While Mg-rich crystalline pyroxene (enstatite) shows emission features at approximately 9.3, 10.6, and 11.6 μm (Chihara et al. 2002), Mg-rich crystalline olivine (forsterite) has a strong peak feature at approximately 11.2 μm, as well as peaks at 10.0 and 11.9 μm (Koike et al. 2003).

To determine the physical properties of the dust grains in comet 21P/G-Z, we used the thermal emission model for cometary dust grains (Ootsubo et al. 2007a). This model is similar to that considered by Harker et al. (2002, 2007) but with some updated optical constants and is often used to reproduce the observed spectrum and to derive the physical properties of dust grains in comets (Ootsubo et al. 2007a, 2007b; Shinnaka et al. 2018). The model incorporates five types of minerals: amorphous with olivine composition and pyroxene composition (Mg:Fe = 1:1; Dorschner et al. 1995), amorphous carbon (Preibisch et al. 1993), and crystalline olivine (Mg-rich; Koike et al. 2003) and crystalline pyroxene (Mg-pure; Chihara et al. 2002). We assumed that all grains in the comet are at the same heliocentric distance (same as the distance of the nucleus from the Sun) and their temperature is in radiative equilibrium with the solar radiation field. We calculated the temperatures of each grain (with a given radius, composition, crystalline or amorphous form, and porosity). The thermal emission from each grain was modeled as the product of the emission efficiency factor and the Planck function for the temperature of the dust. Finally, we integrated the thermal emission from each grain over the size distribution of the cometary grains expressed by the Hanner size distribution (Hanner 1983), represented by $n(a) = (1 - a_0/a)^M (a_0/a)^N$, where $a$ and $a_0$ (= 0.1 μm) are the grain radius and the minimum grain radius in μm, respectively. $N$ and $M$ are power-law indices of dust size distribution and they are related to the peak grain radius $a_p = a_0 (M + N)/N$. Porosity parameter ($D$) ranging from 2.5 to 3.0 ($D$ = 2.5, 2.609, 2.727, 2.857, and 3.0; Ootsubo et al. 2007a) corresponds to the vacuum fraction of filled volume $f = 1 - (a/a_0)^{D-3}$. By comparing the synthesized spectra with the observed spectrum, we determined the mass fraction of each mineral in the cometary grains. Further details of the model are also provided in Ootsubo et al. (2007a) and Shinnaka et al. (2018).

The N-band spectrum of comet 21P/G-Z in the wavelength range from 8.0 to 12.5 μm and the photometric data at 8.8, 12.4, and 18.8 μm were fitted with the model spectrum by the $\chi^2$ minimization technique with the Marquardt-Levenberg algorithm. The wavelength region from 9.4 to 9.8 μm was not used for the fitting to avoid the influence of the strong telluric ozone ($O_3$) absorption band. The best-fit synthetic spectrum is shown with the observed N-band spectrum of comet 21P/G-Z in Figure 1. As demonstrated in the figure, the modeled spectrum fit of the data failed to reproduce the observed spectrum (a reduced-$\chi^2$ is 2.07). The photometric data at 18.8 μm, which is even a single data point but important to test the soundness of the model fit, is, in particular, not reproduced by the model. The 9.2 μm feature can be partially reproduced by the crystalline pyroxene, but not entirely. More material(s) is obviously needed to reproduce the emission peaks at 8.2, 8.5, and 9.2 μm. Those features have never been observed in previous mid-infrared observations of other comets. Some features that originate from PAHs or hydrogenated amorphous carbons (HACs) might contribute to those unidentified infrared (UIR) features (Draine & Li 2007). For example, PAH molecules are expected to exhibit the peaks around 8.2, 8.5, 11.2, and 12.7 μm. The relative strengths of those features depend not only on the ionization degree of PAHs but also on the size of PAHs (e.g., Draine & Li 2007; Shannon & Boersma 2019). Draine & Li (2007) showed the ratios among ~8.3, ~8.6, ~11.3 and ~12.7 μm features of PAHs are ~1:~4:~1.4:~0.7 and ~1:~4:~10:~5 for ionized and neutral PAHs, respectively. The observed spectrum seems consistent with the former case rather than the latter case.

Lisse et al. (2006, 2007) claimed the detections of PAH features in the mid-infrared spectra of comets 9P/Tempel 1 and C/1995 O1 (Hale-Bopp), while the detection of the PAH features in a comet is still under debate (Crovisier & Bockelée-Morvan 2008). Furthermore, we need additional material to explain the unidentified infrared (UIR) emission at ~9.2 μm even if we consider the contribution of PAH molecules. Therefore, we fitted the spectrum by masking the wavelength regions of 8.1–8.6 μm, 9.0–9.4 μm, 11.0–11.4 μm, and 12.6–12.8 μm (Figure 2). The wavelength regions, except ~9.2 μm, seemed to overlap with the emission wavelengths of PAHs or HACs. The residual spectrum obtained after the best-fit model spectrum was subtracted from the observed spectrum is shown in Figure 3. Some of the residual peaks found in Figure 3 were fitted by a linear combination of Gaussian functions to derive both the central wavelength and full width at half maximum (FWHM) of the four features (at ~8.2, ~8.5, ~9.2, and ~11.2 μm). Table 2 lists the central wavelengths and FWHMs of the features. The central wavelengths of features (A), (B) and (D) (~8.23, ~8.47, and ~11.17 μm) are close to the wavelengths of PAHs (at 8.3, 8.6, and 11.2 or 11.3 μm, respectively) but slightly different (Draine & Li 2007). Furthermore, no corresponding feature is found for the feature (C) at ~9.2 μm in PAHs usually observed in astronomical objects.

Note that the 11.2 μm feature (D) is usually attributed to crystalline silicate (e.g., Mg-rich olivine, forsterite). Because we cannot distinguish the emission features from those of PAHs at ~11.3 μm and crystalline olivine at ~11.2 μm, we eliminated the wavelength region around ~11.2 μm for the fitting in Figure 2. Therefore, we may underestimate the crystalline olivine grains in Figure 3. If we include the wavelength region around the 11.2 μm feature for the fitting, the best-fit result and its residual are shown in Figures 4 and 5, respectively. The residuals shown in Figure 3 and Figure 5 are only different for the peak at ~11.2 μm. Table 3 also lists the central wavelengths and FWHMs of the features in this case. However, the absence of strong UIR feature at ~11.3 μm, which is attributed to PAHs, in the observed spectrum is unlikely despite the presence of the UIR feature at ~8.5 μm, which is also attributed to PAHs.

Table 4 summarizes the physical parameters of dust grains in comet 21P/G-Z based on the fittings discussed here. In any cases, we found the presence of strong features (A), (B), and (C) even though we assumed that the 11.2 μm peak (D) entirely originated from crystalline olivine. On the other hand, the crystalline fraction of silicate grain is quite hard to derive accurately. If we eliminate the 11.2 μm peak for the fitting (Figure 2), a mass fraction of crystalline components in silicates ($f_{cry}$) is derived as $f_{cry}$ = 0.45 +0.33/–0.11, which is consistent with $f_{cry}$ in other comets (from ~0.3 to ~0.8, in Table 3 by Shinnaka et al. 2018).

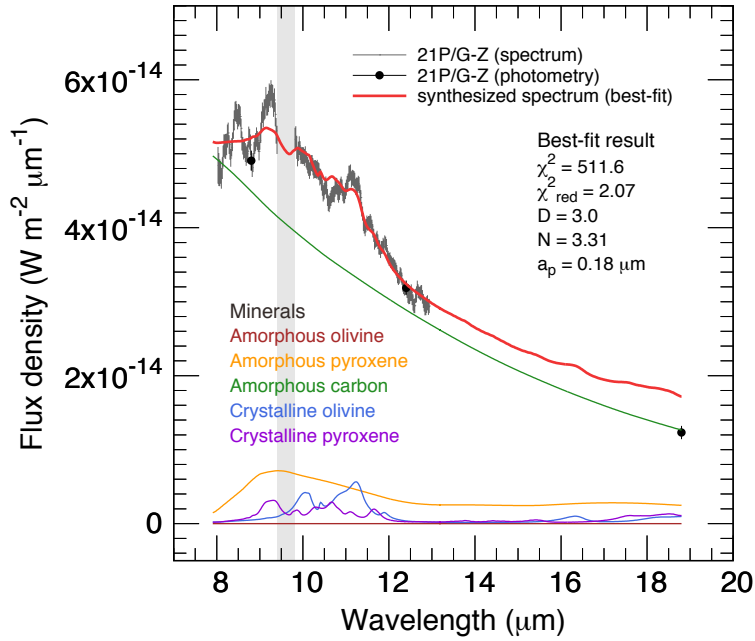

**Figure 1**: Best-fit thermal emission model for comet 21P/G-Z for fitting with all wavelength regions (8–18.8 μm). The data at 9.4–9.8 μm affected by the strong absorption of telluric ozone are not used for the fitting. The best-fit spectrum cannot reproduce the observed spectrum of 21P (especially, the emission peaks at 8.2, 8.5, and 9.2 μm and the photometric point at 18.8 μm).

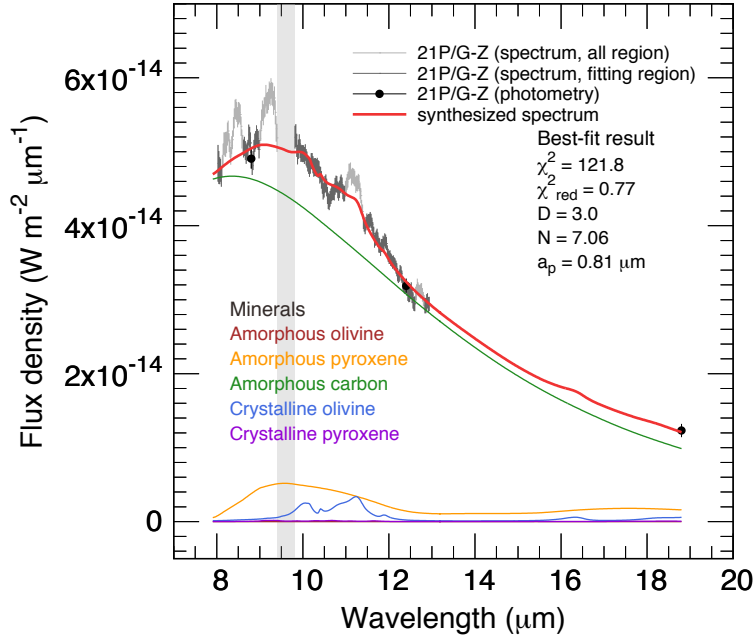

**Figure 2**: Best-fit thermal emission model for comet 21P/G-Z for fitting without PAH features (8.2, 8.5, 11.2, and 12.7 μm) and an unidentified emission (9.2 μm). The data not used in the fit are depicted in gray.

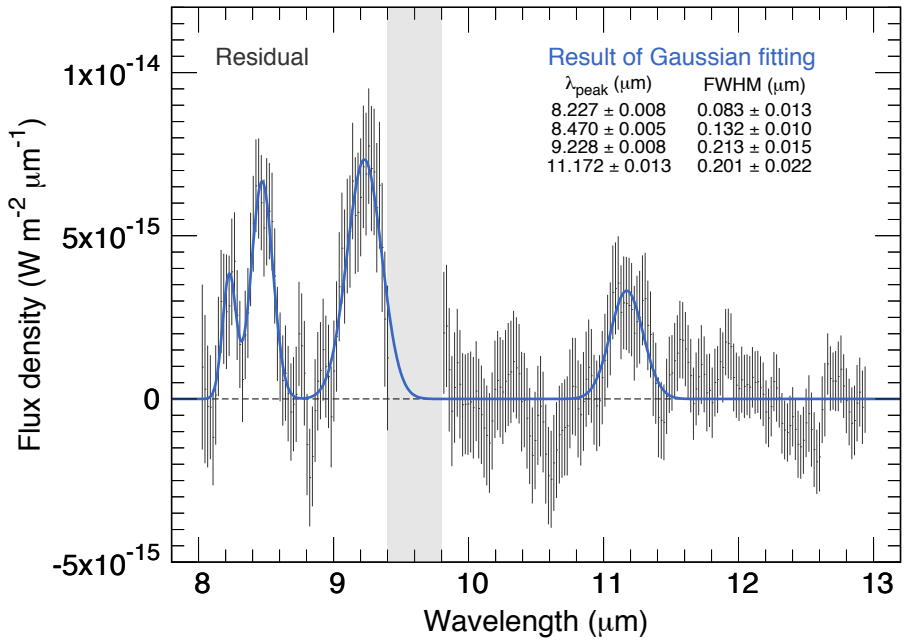

**Figure 3**: Residual spectrum after subtraction of the best-fit model spectrum in Figure 2 from the original spectrum.

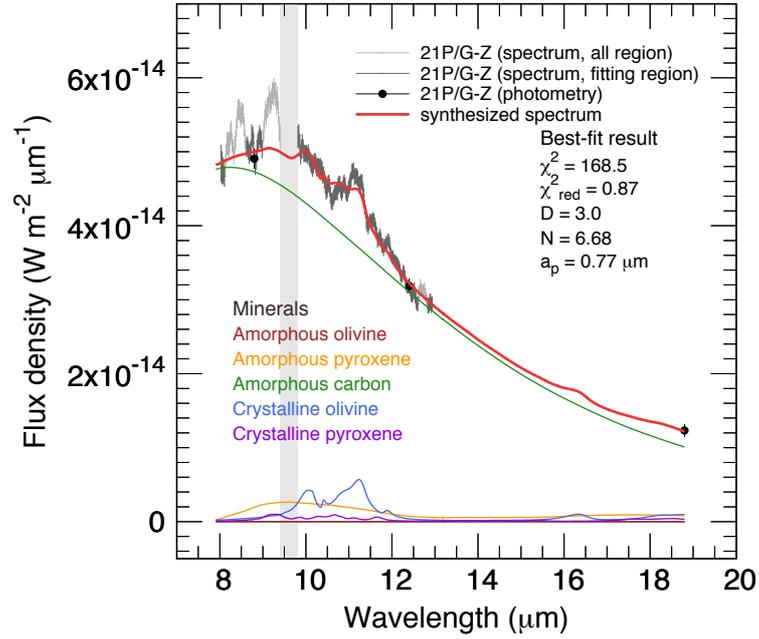

**Figure 4**: Best-fit thermal emission model for comet 21P/G-Z for fitting without possible PAH features (8.2, 8.5, and 12.7 μm) and an unidentified emission (9.2 μm). The data not used in the fit are depicted in gray.

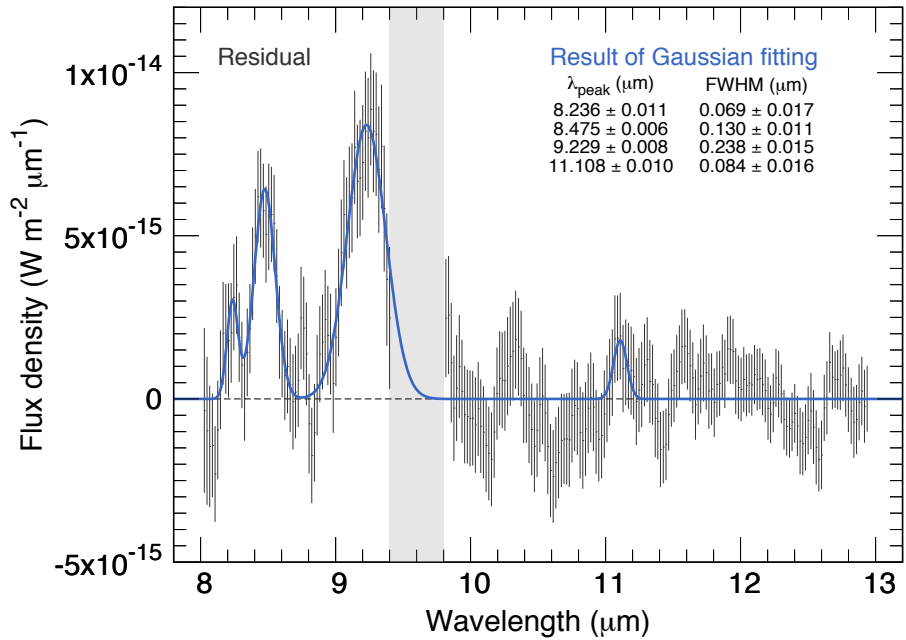

**Figure 5**: Residual spectrum after subtraction of the best-fit model spectrum in Figure 4 from the original spectrum.

Table 1. Summary of Subaru+COMICS Observation of comet 21P/G-Z.

| UT Date | Time | Object | Integration Time (s) | Observing mode | Airmass |
|---|---|---|---|---|---|
| 2005 July 5 | 15:08-15:11 | 21P/G-Z | 121.1 | N8.8 | 1.74 |
| | 15:15-15:18 | | 121.5 | N12.4 | 1.67 |
| | 15:24-15:28 | | 121.3 | Q18.8 | 1.60 |
| | 15:33-15:50 | | 361.6 | NL Spec | 1.49 |
| 2005 July 5 | 13:49 | HD3712 | 10.8 | N8.8 | 1.41 |
| | 13:51 | | 10.6 | N12.4 | 1.40 |
| | 14:15 | | 10.8 | Q18.8 | 1.35 |
| | 14:22 | | 45.2 | NL Spec | 1.34 |

Table 2: Unidentified features found in the spectrum of comet 21P/G-Z from Figure 3.

| ID | Central wavelength (μm) | FWHM (μm) |
|---|---|---|
| A | 8.227 ± 0.008 | 0.083 ± 0.013 |
| B | 8.470 ± 0.005 | 0.132 ± 0.010 |
| C | 9.228 ± 0.008 | 0.213 ± 0.015 |
| D | 11.172 ± 0.013 | 0.201 ± 0.022 |

Table 3: Unidentified features found in the spectrum of comet 21P/G-Z from Figure 5.

| ID | Central wavelength (μm) | FWHM (μm) |
|---|---|---|
| A | 8.236 ± 0.011 | 0.069 ± 0.017 |
| B | 8.475 ± 0.006 | 0.130 ± 0.011 |
| C | 9.229 ± 0.008 | 0.238 ± 0.015 |
| D | 11.108 ± 0.010 | 0.084 ± 0.016 |

**Table 4:** Derived physical parameters of dust grains in comet 21P/G-Z.

| Case (†) | Mass fraction of dust grains | | | | | Dust properties | | | $\chi^2$ (DoF) for the best-fit | $\chi^2_{red}$ |
|---|---|---|---|---|---|---|---|---|---|---|
| | Amorphous | | | Crystalline | | $D$ | $a_p$ (μm) | $N$ | | |
| | Olivine | Pyroxene | Carbon | Olivine | Pyroxene | | | | | |
| Case 1 | 0.0 | $0.13^{+0.05}/_{-0.11}$ | $0.61^{+0.11}/_{-0.19}$ | $0.08^{+0.35}/_{-0.02}$ | $0.18^{+0.04}/_{-0.08}$ | 3.0 | $0.18^{+0.32}/_{-0.08}$ | $3.31^{+1.66}/_{-0.29}$ | 511.6 (247) | 2.07 |
| Case 2 | $0.00^{+0.03}/_{-0.00}$ | $0.12^{+0.07}/_{-0.04}$ | $0.78^{+0.08}/_{-0.09}$ | $0.08^{+0.06}/_{-0.02}$ | $0.02^{+0.10}/_{-0.02}$ | 3.0 | $0.81^{+0.23}/_{-0.27}$ | $7.06^{+5.77}/_{-2.35}$ | 121.8 (158) | 0.77 |
| Case 3 | 0.0 | $0.04^{+0.08}/_{-0.04}$ | $0.77^{+0.54}/_{-0.72}$ | $0.15^{+0.08}/_{-0.03}$ | $0.04^{+0.05}/_{-0.04}$ | 3.0 | $0.77^{+0.25}/_{-0.31}$ | $6.68^{+5.57}/_{-2.36}$ | 168.5 (194) | 0.87 |

(†) Case 1: by masking telluric ozone region only (see Figure 1).

Case 2: by masking the wavelength regions of 8.1–8.6 μm, 9.0–9.4 μm, 11.0–11.4 μm, and 12.6–12.8 μm + telluric ozone region (see Figure 2).

Case 3: by masking the wavelength regions of 8.1–8.6 μm, 9.0–9.4 μm, and 12.6–12.8 μm + telluric ozone region (see Figure 4).

(*) $D$: a porosity parameter of dust grains, $a_p$: a peak size of dust size distribution, $N$: power index of the distribution function (Hanner 1983). Although we calculated modeled spectra for $D$ ranging from 2.5 to 3.0 ($D$ = 2.5, 2.609, 2.727, 2.857, and 3.0), $D$ = 3.0 is the best to reproduce the observed spectrum. Error-bars are representative of the 68%-confidence intervals.

4. Discussion

The UIR features found in the mid-infrared spectra of classical novae are similar to the emission features in the spectrum of 21P/G-Z. Classical novae-forming dust grains sometimes exhibit UIR features at wavelengths slightly different from those of PAHs, e.g., in V705 Cas (Evans et al. 2005). Such a difference could be explained by contamination by N- or O-atoms from PAHs. The strongest feature (C) in the spectrum of comet 21P/G-Z was also observed in the spectrum of nova V705 Cas at 9.20 μm (Evans et al. 2005). Several emission features that are not typically present in "normal" UIR carriers can also be explained by N- and O-atom contamination. Although the carrier of the ~9.2 μm UIR feature is still unknown, Evans et al. (2005) suggested that significant nitrogen contamination was responsible for this feature in the case of novae. The carrier of the ~9.2 μm feature observed in the spectrum of comet 21P/G-Z may be some organic molecules contaminated with nitrogen, as demonstrated in the laboratory (Grishko & Duley 2002).

Many laboratory studies have searched for the carriers of UIR features (Molpeceres et al. 2017, and references therein). The spectral features in the 9–9.5 μm region are reported in laboratory studies. For example, those features are found in the IR absorbance spectra of some organic molecules produced by vacuum pyrolysis of pure cellulose at temperatures up to ~600 K and ~730 K (Blanco et al. 1988). The features at approximately 9–9.5 μm were seen only for the materials heated to ~600 K. The organic molecules formed in this case were considered a mixture of aliphatic and aromatic hydrocarbons. Both the 3.4 μm feature from aliphatic hydrocarbons and the 11.4 μm feature from aromatic hydrocarbons were strong. However, if cellulose was heated up to ~730 K, the 3.4 μm feature was very weak (i.e., the resultant organic molecules were mainly aromatic hydrocarbons), and the features at approximately 9–9.5 μm were not observed. This experimental result implies that the ~9.2 μm feature observed in comet 21P/G-Z can be attributed to aliphatic hydrocarbons, which cannot withstand temperatures higher than ~730 K. This fact shed light on the origin of the organic materials expected due to the presence of some UIR features in comet 21P/G-Z. On the other hand, it is reported that the feature around 9.3 μm appears for PAHs with a smaller number of carbon atoms (≤ 30 carbon atoms) based on recent theoretical studies (Shannon & Boersma 2019). However, if carriers of the ~9.2 μm feature detected in comet 21P/G-Z are smaller-sized PAHs, a strong emission feature at ~12 μm should also appear (Shannon 2019, private comm.). No clear detection of ~12 μm feature in our spectrum also supports the hypothesis that the carriers of the ~9.2 μm feature are aliphatic hydrocarbons.

The presence of the strong UIR features in the mid-infrared spectra of comets is quite rare; such features have never been observed in previous studies. Therefore, the origin of comet 21P/G-Z seems to be unusual. As demonstrated by near-infrared spectroscopy (DiSanti et al. 2013), simple organic molecules, such as $C_2H_6$ and $CH_3OH$, that are observed in the 3 μm region were depleted in the spectrum of comet 21P/G-Z. More complex organic molecules, however, might be present in the comet, as shown here. This result is consistent with the optical polarimetric observations of comet 21P/G-Z (Kiselev et al. 2000a, 2000b). The comet is depleted in simple organic molecules but enriched in complex organic molecules.

Figure 6 demonstrates the peculiarity of comet 21P/G-Z in blackbody normalized mid-infrared spectra of comets. The spectra of comets 73P/Schwassmann-Wachmann 3 fragments of B and C, 9P/Tempel 1 just after the Deep Impact, 17P/Holmes during its mega-outburst, and C/2002 V1 (NEAT) are plotted with 21P/G-Z (Shinnaka et al. 2018; Ootsubo et al. 2007b; Honda et al. 2004). Comets 21P/G-Z and 73P/Schwassmann-Wachmann 3 are classified into the same taxonomic group "G-Z type" (Fink 2009), and these comets also showed the negative wavelength gradient of linear polarization for dust grains (Kiselev et al. 2015). However, the mid-infrared spectra of the fragments B and C of comet 73P/Schwassmann-Wachmann 3 are quite different from that of comet 21P/G-Z as shown in Figure 6. Although the entire strength of 10 μm silicate feature of comet 21P/G-Z is comparable to those of 73P/Schwassmann-Wachmann 3 and weaker than other comets, the 8.2 and 8.5 μm features are quite strong in comet 21P/G-Z compared to other comets. The 9.2 μm feature may also exist in comet 17P/Holmes (in the shoulder of a broad 10 μm silicate feature of the comet). Moreover, comet 21P/G-Z probably shows the PAH feature at 12.7 μm but not clear for other comets. Table 5 lists the physical properties of dust grains in comets 21P/G-Z, 73P/Schwassmann-Wachmann 3 fragments B and C, 17P/Holmes (Jupiter-family comets), C/2002 V1 (NEAT) and C/2001 Q4 (NEAT) (Oort Cloud comets) (Ootsubo et al. 2007a; Harker et al. 2011, 2017; Shinnaka et al. 2018). The crystalline fraction of silicate grains in comet 21P/G-Z is moderate among comets, and the peak size of dust grains in comet 21P/G-Z is larger than those in other comets.

The enrichment of complex organic molecules is probably due to gas-phase chemistry at the warm temperatures (> 100 K) in the solar nebula, or due to the cold-temperature chemistry on the grain surface at ~10 K in a natal dark cloud followed by the photolysis in ice mantle by high energetic particles in the solar nebula. The birthplace of comet 21P/G-Z in the solar nebula might be closer to the Sun than the birthplace of other comets (usually considered to be ~5–30 au from the Sun). However, the mass fraction of the crystalline component in silicate ($f_{cry}$) is 0.45 for comet 21P/G-Z and is normal compared to other comets (Table 5). Because the crystalline silicate grains were considered to be formed and transported from the inner high-temperature region of the early solar nebula, comets formed at further distances from the Sun might contain less crystalline silicate grains compared to amorphous silicate grains originating from interstellar dust grains. The similarity of $f_{cry}$ between comet 21P/G-Z and other comets indicates that comet 21P/G-Z formed at a similar distance from the Sun in the solar nebula as other comets formed (Shinnaka et al. 2018).

We propose that comet 21P/G-Z formed in the circumplanetary disk around a gas giant planet (Jupiter or Saturn), where the distance from the Sun is similar to that of the birthplaces of other comets but the temperature is warmer. The rocky/icy regular satellites (in their prograde orbits) around gas giant planets such as Jupiter and Saturn are believed to have formed in gas disks around the planets, called circumplanetary disks, similar to the solar nebula around the Sun (Lunine & Stevenson 1982; Shibaike et al. 2017). A circumplanetary disk is formed by gas accretion from the solar nebula to the gas giant planet as the planet forms. The temperature of a circumplanetary disk strongly depends on the mass accretion flux to the disk from the ambient gas in the solar nebula where the gas temperature was < 100 K in the forming region of the gas giant planets (Estrada et al. 2009; Tanigawa et al. 2012). The temperature of the circumplanetary disk is expected to be too high for ice grains to exist (> ~150 K as the snow line) when the mass accretion flux is very high, corresponding to the initial stage of the planet formation. However, as the gas giant planet grows, the gap that is a low-density annulus region near the planet orbit formed in the solar nebula and the mass accretion flux to the circumplanetary disk decreased; i.e., the temperature of circumplanetary disk became low enough for ice to exist. During the final stage of the satellite formation, many icy satellitesimals (small bodies) are present as building blocks of icy satellites in the disk. As demonstrated by previous studies about the chemical evolution of interstellar media, various kinds of complex organic molecules have been observed in warm environments (> 100 K), such as hot molecular cores (e.g., Choudhury et al. 2015; Bergner et al. 2017; Bianchi et al. 2019). Therefore, complex organic molecules formed by gas-phase chemical reactions under warm conditions (in the earlier evolutional stage of a circumplanetary disk) might finally be frozen and incorporated into satellitesimals as the circumplanetary disk cools down. Furthermore, the catalytic conversion of methanol to larger organic molecules over heated crystalline silicates grains (up to ~700 K) demonstrated by Li et al. (2018) might occur in a circumplanetary disk. Because the equatorial plane of a circumplanetary disk is nearly the same as the equatorial plane of a solar nebula, the icy satellitesimals thrown out from a circumplanetary disk may be a source of "chemically unique" Jupiter-family comets, such as 21P/G-Z.

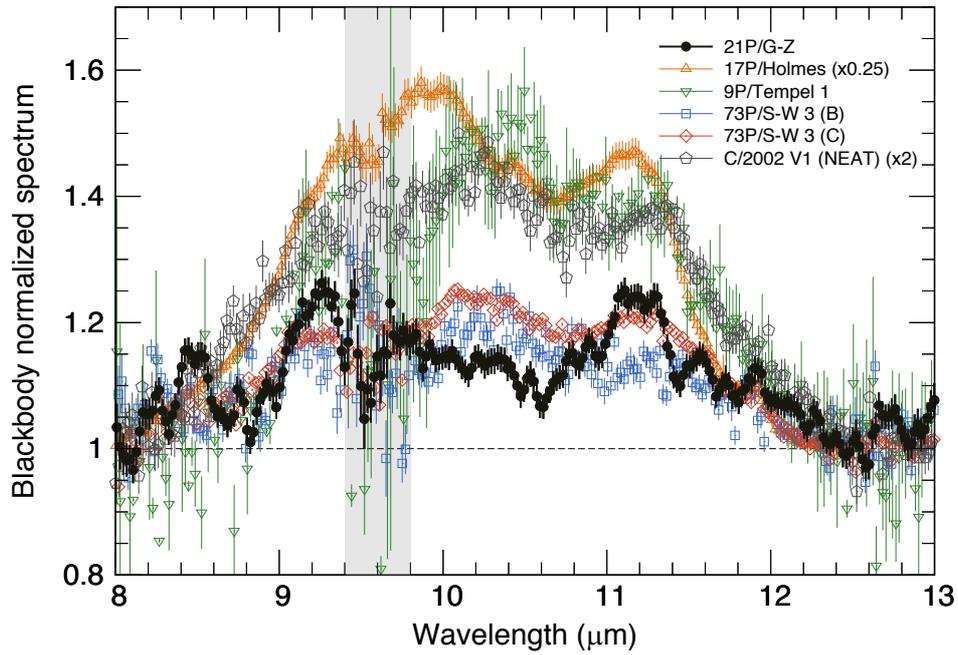

**Figure 6**: Blackbody normalized comet spectra. The spectra of comets 73P/Schwassmann-Wachmann 3 (fragments B and C), 9P/Tempel 1 (just after the Deep Impact), 17P/Holmes (on October 25 during its mega-outburst), and C/2002 V1 (NEAT) are plotted with 21P/G-Z (Honda et al. 2004; Ootsubo et al. 2007b; Shinnaka et al. 2018).

**Table 5:** Physical parameters of dust grains in comets.

| Comet (†) | Mass fraction of dust grains | | | | | Dust properties | | | |
|---|---|---|---|---|---|---|---|---|---|
| | Amorphous | | | Crystalline | | $D$ | $a_p$ (μm) | $N$ | $f_{cry}$ (*) |
| | Olivine | Pyroxene | Carbon | Olivine | Pyroxene | | | | |
| 17P/Holmes | 0.31 | 0.32 | 0.12 | 0.18 | 0.08 | 3.00 | 0.1 | 3.3 | 0.29 |
| 21P/G-Z | 0.00 | 0.12 | 0.78 | 0.08 | 0.02 | 3.00 | 0.8 | 7.1 | 0.45 |
| 73P-B/S-W | 0.04 | 0.19 | 0.43 | 0.07 | 0.27 | 2.73 | 0.5 | 3.4 | 0.59 |
| 73P-C/S-W | 0.14 | 0.00 | 0.60 | 0.26 | 0.00 | 2.73 | 0.3 | 3.4 | 0.65 |
| C/2002 V1 | 0.05 | 0.15 | 0.42 | 0.37 | 0.01 | 2.86 | 0.5 | 3.5 | 0.66 |
| C/2001 Q4 | 0.13 | 0.05 | 0.37 | 0.42 | 0.03 | 3.00 | 0.2 | 3.6 | 0.71 |

(†) Values are taken from Shinnaka et al. (2018) for comet 17P/Holmes on October 25, from Table 4 (case 2) for comet 21P/G-Z, from Harker et al. (2011, 2017) for the fragments B and C of comet 73P/Schwassmann-Wachmann 3, and from Ootsubo et al. (2007a) for comets C/2001 Q4 (NEAT) and C/2002 V1 (NEAT).

(*) Mass fraction of sub-μm crystalline silicate grains relative to the total sub-μm silicate grains (amorphous + crystalline).

5. Summary

We performed N-band low-resolution spectroscopic observations ($\lambda$ = 8–13 μm, $\lambda/\Delta\lambda \sim 250$) and narrow-band photometric observations (at $\lambda$ = 8.8, 12.4, and 18.8 μm with $\Delta\lambda$ = 0.8, 1.2, and 0.9 μm, respectively) of comet 21P/G-Z on UT 2005 July 5. The obtained spectrum shows the unidentified infrared emission features at ~8.2 μm, ~8.5 μm, ~9.2 μm, and ~11.2 μm. The features at ~8.2 μm, ~8.5 μm, and ~11.2 μm could be attributed to PAHs (or HACs) contaminated by N- or O-atoms, although part of the feature at ~11.2 μm comes from crystalline olivine. The feature at ~9.2 μm might originate from aliphatic hydrocarbons. Thus, the parent comet of the October Draconids meteor shower, comet 21P/G-Z, is enriched in complex organic molecules that may be pre-biotic. The presence of complex organic molecules in the dust grains of comet 21P/G-Z indicates that comets similar to comet 21P/G-Z probably originated in circumplanetary disks in a solar nebula and might have delivered pre-biotic organic molecules to ancient Earth via meteoroids (or by the impact of those comets).

Acknowledgements

This paper is based on data collected by the Subaru Telescope, which is operated by the National Astronomical Observatory of Japan. This study is financially supported by MEXT Supported Program for the Strategic Research Foundation at Private Universities, 2014–2018 (No. S1411028). T.O. is supported by JSPS KAKENHI Grant-in-Aid for Scientific Research (C) JP17K05381 and (A) JP19H00725.

Data availability

Raw data of 21P/G-Z, HD3712, flat, and dark frames can be retrieved from the Subaru SMOKA Science Archive (https://smoka.nao.ac.jp).